\documentclass[preprint,aps,tightenlines,showkeys,nofootinbib,superscriptaddress]{revtex4-2}
\usepackage{color}
\usepackage{enumerate}
\usepackage{latexsym}
\usepackage{graphicx}
\def\be{\begin{equation}}
\def\ee{\end{equation}}
\def\bea{\begin{eqnarray}}
\def\eea{\end{eqnarray}}

\begin{document}

%\preprint{arXiv:2402.xxxxx}
\today\\

\title{Deflection of light by a compact object with electric charge and magnetic dipole in Einstein-Born-Infeld gravity}

\author{Jin Young Kim \footnote{E-mail address: jykim@kunsan.ac.kr} }
\affiliation{Department of Physics, Kunsan National University,
Kunsan 54150, Korea}

\begin{abstract}

We consider the bending of light around a compact astrophysical object with both the electric field and the magnetic field in Einstein-Born-Infeld theory. 
From the null geodesic of a light ray passing a massive object with electric charge and magnetic dipole, the effective metric was obtained from the light cone condition reflecting the nonlinear electromagnetic effects. 
We found the asymptotic form of the effective metric up to the first order in gravitational constant $G$ and Born-Infeld parameter $1/\beta^2$ on the equatorial plane. Then we compute the bending angle of light from the geodesic equation. 
The result includes particular cases where only one type of field is present taking the appropriate limits. 

\end{abstract}

%\pacs{42.25.Dd, 12.20.-m, 12.90.+b }


\maketitle

\newpage

\section{Introduction}

Deflection of light by a massive object is one of the important result of general relativity. It is the starting point of gravitational lensing theory. 
The shadow image of black hole observed by the Event Horizon Telescope has become an important topic to study the black hole physics \cite{EHT1, EHT2, EHT3}. In general relativity, any physical quantity with energy and momentum can affect the spacetime structure and can cause the bending of light. When a light ray is passing around a massive object with electomagnetic field, the electromagnetic field can also affect the bending angle. The bending angle can be computed from the geodesic equation using the metric obtained from the Einstein-Maxwell equation. 

Astronomical objects with extremely strong electromagnetic field attract interest. The magnetic field on the surface of a magnetar is estimated up to the order of $10^{11} {\rm T}$ \cite{duncan,thompson}. This field strength is the estimated lower bound of the Born-Infeld parameter \cite{Jackson} where nonlinear electromagnetic effects can be considered. When the electric or magnetic field made by a compact astrophysical object is strong enough, one can consider the bending of light accommodating the nonlinear electromagnetic effects. In  Einstein-Maxwell theory, the presence of electromagnetic field affects only the energy-momentum tensor and the electromagnetic field equations are linear. So the null geodesic of the electromagnetic wave is the same as the null geodesic of the gravitational wave. However, in Einstein-Born-Infeld theory, the null geodesic of the electromagnetic wave is not the same as that of the gravitational wave due to the nonlinear electromagnetic effects \cite{Plebanski}. 

In the generalized Born-Infeld-type nonlinear electrodynamics, one can compute weak bending angles of light by a charged black hole \cite{kim2021} and a magnetar \cite{kim2022} using the effective null geodesic. These computations correspond to the cases where only one type of electromagnetic field (purely electric or purely magnetic) is present. Recently it is observed that black holes can have magnetic field by the Event Horizon Telescope \cite{EHT2021}. It seems of interest to consider the bending of light by compact astrophysical objects which have both the electric field and the magnetic field. For example, one can compute the bending angle of light by a charged black hole with magnetic field or a magnetar with electric charge in Einstein-Born-Infeld theory.

The organization of the paper is as follows. In Sec. II, we consider the geometrical aspects of a massive object with both the electric charge and the magnetic dipole moment. We outline how to obtain the effective metric and the electromagnetic four-potential in Einstein-Born-Infeld theory. We found the static axially symmetric solution of the effective metric on the equatorial plane of a magnetic dipole. In Sec. III, using the geodesic equation on the equatorial plane,
we compute the bending angle of light up to the first order in gravitational constant $G$ and Born-Infeld parameter $1/\beta^2$. 
We show that the result includes particular cases where only one type of field is present.  In Sec. IV, we summarize and discuss.

\section{Effective null geodesic of the electromagnetic wave}

The Einstein-Born-Infeld action is described by \cite{Born,BornInfeld}
\be 
S = \int d^4 x \sqrt{-g} \left ( \frac{R}{16 \pi } +  {\cal L} \right ) , \label{EBIaction}
\ee
where $ {\cal L}$ is 
\be
 {\cal L} = \beta^2 \left ( 1- \sqrt{ 1 + \frac{2 S}{\beta^2} - \frac{P^2}{\beta^4} } \right ) .
 \label{cbilagran}
\ee
Here $\beta$ is the classical Born-Infeld parameter characterizing the possible maximum value of the field strength, $S$ and $P$ are Lorentz-invariants defined by
\be
 S = \frac{1}{4} F_{\mu \nu} F^{\mu \nu} = \frac{1}{2} ( {\bf B}^2 -  {\bf E}^2 )  , 
 ~~~ P = \frac{1}{4} F_{\mu \nu} {\tilde F}^{\mu \nu} =  {\bf E} \cdot {\bf B} ,
 \label{defSP}
\ee
$F_{\mu \nu} = \partial_\mu A_\nu - \partial_\nu A_\mu $ is the field strength tensor,  
and ${\tilde F}_{\mu \nu}  = \frac{1}{2} \epsilon_{\mu \nu \alpha \beta} F^{\alpha \beta} $ is the dual tensor. 
 We use the unit system with $1 / 4 \pi \epsilon_0 = \mu_0 / 4 \pi = c = 1$. However, we keep the gravitational constant $G$ and use it as the perturbation parameter. 

The equations of motion are obtained by varying the action with respect to $g_{\mu \nu}$ and $A_\mu$ with the Bianchi identity
\be
 R_{\mu \nu} - \frac{1}{2} g_{\mu \nu} R = 8 \pi G T_{\mu \nu} ,  \label{Eineq}
\ee
\be
 \nabla_\mu \left [ \frac{1} {  \sqrt{ 1 + \frac{2 S}{\beta^2} - \frac{P^2}{\beta^4} } }
\left ( F^{\mu \nu} - \frac{ P}{ \beta^2 } {\tilde F}^{\mu \nu} \right ) \right ]=  0,  \label{Maxeq}
\ee
where $\nabla_\mu$ is the covariant derivative and $T_{\mu \nu}$ is the energy momentum tensor given by
\be 
T_{\mu \nu} =  \beta^2 \left ( 1 - \sqrt{ 1 + \frac{2S}{\beta^2}  - \frac{ P^2 }{\beta^4 } } \right ) g_{\mu \nu}
  +  \frac{1} {  \sqrt{ 1 + \frac{2 S}{\beta^2} - \frac{P^2}{\beta^4} } } F_\mu^{~\rho}
\left ( F_{ \rho \nu} - \frac{ P}{ \beta^2 } {\tilde F}_{ \rho \nu} \right )  .
\ee
In the limit $\beta \to \infty$, the above equations reduce to the Einstein-Maxwell equations
\be
 R_{\mu \nu} - \frac{1}{2} g_{\mu \nu} R 
= 8 \pi G \left ( F_\mu^{~\rho} F_{ \rho \nu} 
 - \frac{1}{4}  F_{\rho \sigma} F^{\rho \sigma}  g_{\mu \nu} \right ),   \label{eineqlin}
\ee
\be
 \nabla_\mu  F^{\mu \nu} =  0.  \label{maxeqlin}
\ee

We want to find the metric tensor and electromagnetic potential by a compact object having both the electric charge and the magnetic dipole. Because of the axial symmetry, we use the cylindrical coordinates, $x^\mu = ( t, r, z , \phi)$, taking the direction of magnetic dipole as $z$-axis. When a compact object has only one type of electromagnetic field, the metric tensor has only diagonal elements because the Poynting vector is zero. When it has both the electric and the magnetic fields, there are off-diagonal elements in the energy-momentum tensor so that the metric tensor also has off-diagonal elements. We take the following form of the metric tensor and the electromgnetic four-potential 
\be
g_{\mu \nu} = \pmatrix { e^\rho  & 0  & 0  & \omega \cr 
0 & -e^\lambda  &  0 & 0 \cr  
0&  0& -e^\lambda &0  \cr
\omega&  0& 0  & - r^2 e^{-\nu}  } ,  \label{metricftn}
\ee
\be
A_\mu = ( \Phi, 0, 0, - \psi ) , \label{four-pot}
\ee
 where $\rho, \lambda, \nu, \omega, \Phi$, and $\psi$  are functions of $r$ and $z$ only. 

Substituting Eqs. (\ref{metricftn}) and (\ref{four-pot}) into Eqs. (\ref{Eineq}) and (\ref{Maxeq}), we have coupled nonlinear differential equations of six unknown functions, $\rho, \lambda, \nu, \omega, \Phi$, and $\psi$. 
The boundary conditions for the metric and the electromagnetic four-potential should be as follows. At infinity, the metric functions become asymptotically flat and the electromagnetic field should vanish. In the limit where the electric charge and the magnetic dipole moment is zero, the metric should approach the Schwarzschild metric. In the limit where the mass is zero and $\beta \rightarrow \infty$, the metric is flat everywhere and the potentials in Eq. (\ref{four-pot}) should be the four-potential in flat space given by
\be
 \Phi = \frac{q}{ \bar R}, ~~~\psi = \mu \frac{r^2}{{\bar R}^3}  , \label{flatfour-pot}
\ee
where $q$ is the Coulomb charge, $\mu$ is the magnetic dipole moment, and we denote the spherical distance ${\bar R}= \sqrt{r^2 + z^2}$ to avoid the confusion with the scalar curvature. 
The electromagnetic field tensor can be written as
\be
F_{\mu \nu} = \partial_\mu A_\nu -\partial_\nu A_\mu 
 = \pmatrix {0  & -\Phi_r & -\Phi_z  & 0 \cr 
\Phi_r & 0  &  0 & -\psi_r \cr  
\Phi_z&  0& 0 &-\psi_z  \cr
0&  \psi_r& \psi_z & 0  } ,  \label{Fco}
\ee
where the subscripts $r$ and $z$ denote the partial derivatives. 

Finding the complete analytic solution of Eqs. (\ref{metricftn}) and (\ref{four-pot}) is not easy. However, in Einstein-Maxwell theory, the perturbative solution in powers of $G$ was obtained by Martin and Prechett \cite{martinpritchett} from Eqs. (\ref{eineqlin}) and  (\ref{maxeqlin}). 
Up to the first order in $G$, the series solutions are
\bea
e^\rho &=& 1 - \frac {2GM} {\bar R} + \frac{Gq^2 }{ {\bar R}^2} + \frac {G \mu^2 z^2} {{\bar R}^6} ,    \label{erho}   \\ 
e^\lambda &=& 1 + \frac {2GM} {\bar R}- \frac {Gq^2 z^2}  {{\bar R}^2} - \frac {G \mu^2 (r^4  - 6 r^2 z^2 + 2 z^4 )} {2{\bar R}^8} ,  
                    \label{elambda} \\
e^{- \nu} &=& 1 + \frac {2GM} {\bar R} -\frac { Gq^2 }{ {\bar R}^2} - \frac {G \mu^2 z^2} { {\bar R}^6 },      \label{enu} \\
\omega &=& -\frac { Gq \mu r^2}  {{\bar R}^4} ,     \label{omega}  \\
\Phi &=&  \frac {q}{\bar R} \left [ 1 - \frac { GM} {\bar R} + \frac {Gq^2 }{ 3{\bar R}^2 } - \frac{G \mu^2 (4 r^2  -  z^2 )} { 35{\bar R}^6} \right ] ,      
 \label{Phi} \\
\psi &=& \frac { \mu r^2} { {\bar R}^3} \left [ 1 + \frac {GM} {2{\bar R}} + \frac {Gq^2 }{ 5{\bar R}^2 } \right ].    \label{psi} 
\eea
In Einstein-Maxwell theory, the electromagnetic wave and the gravitational wave follow the same geodesic made by mass, charge, and magnetic dipole. However, in Einstein-Born-Infeld theory, the null geodesic of the electromagnetic wave is different from the null geodesic of the gravitational wave due to the nonlinear coupling with the background electromagnetic field. We will find the perturbative null geodesic of the electromagnetic wave using Eqs. (\ref{erho})-(\ref{psi}).

In Einstein-Born-Infeld gravity, the effective metric for photon can be obtained from the light cone condition \cite{Plebanski, Novello, Delorenci, Breton, Eiroa} as
\be 
\left ( g^{\mu\nu} + \frac {{\cal L}_{SS}} {{\cal L}_S} F^{\mu \alpha} F_\alpha^{~\nu} \right ) k_\mu k_\nu= 0,
\ee
where $g^{\mu\nu}$ is the metric function of the gravitational wave, $k_\mu$ is the four-vector of the electromagnetic wave, ${\cal L}_S = \partial {\cal L} / \partial S$, 
and ${\cal L}_{SS} = \partial^2 {\cal L} / \partial S^2$. Thus one can find the effective metric from 
\be 
 {\tilde g}^{\mu\nu} = g^{\mu\nu} + \frac {{\cal L}_{SS}} {{\cal L}_S} F^{\mu \alpha} F_\alpha^{~\nu} . \label{g_eff}
\ee
Because the classical Born-Infeld parameter is the possible maximum value of the electromagnetic field strength, in the weak field region where the electric field and the magnetic field  are not strong enough compared with $\beta$, one can take
\be 
\frac {{\cal L}_{SS}} {{\cal L}_S} = - \frac{1}{\beta^2} 
\left ( 1 + \frac{ 2S} {\beta^2} - \frac{P^2 } {\beta^4 } \right )^{-1} \rightarrow -  \frac{1} { \beta^2}.
\ee
With this approximation we can compute the effective metric from
\be 
 {\tilde g}^{\mu\nu} = g^{\mu\nu} - \frac{1}{\beta^2}  F^{\mu \alpha} F_\alpha^{~\nu} =  g^{\mu\nu} +  \delta g^{\mu\nu}, \label{gtilde}
\ee
where we defined $ \delta g^{\mu\nu}  \equiv - (1/{\beta^2})  F^{\mu \alpha} F_\alpha^{~\nu} $. 

The details how to find the effective metric are given in Appendix A and the nonzero components of the covariant metric tensor on the equatorial plane ($z=0$) can be written as
\bea
{\tilde g}_{00} &=& 1- \frac{ 2GM} { r}  + \frac{ Gq^2 } {r^2} - \frac{1} { \beta^2 } \frac{ q^2 }{r^4} ,    \\
{\tilde g}_{03} &=& {\tilde g}_{30} = - \frac{Gq \mu}{r^2} + \frac{ 1 }{ \beta^2}\frac{q \mu }{r^4 },   \\
{\tilde g}_{11} &=& - \left ( 1 + \frac{ 2GM }{ r} - \frac{G \mu^2 }{2 r^4} - \frac{ 1} { \beta^2} \frac{ q^2 }{ r^4 }
+ \frac{ 1 }{\beta^2} \frac{  \mu^2 }{r^6} \right ) ,   \\
{\tilde g}_{22} &=& - \left ( 1 + \frac{ 2GM }{ r } - \frac{G \mu^2}{ 2 r^4 } \right ) , \\
{\tilde g}_{33} &=& - r^2 \left ( 1 +\frac{2GM}{ r} - \frac {G q^2} {r^4 } + \frac{ 1} { \beta^2} \frac{ \mu^2} {r^6 } \right ). 
\eea
In the above metric, terms with $G$ originated from the energy-momentum tensor $T_{\mu\nu}$ and terms with $1 /\beta^2$ originated from the non-linear electrodynamic effect. 
For $q = \mu =0$, the above metric reduces to the Schwarzschild metric. For $\mu =0$, it reduces to the leading-order Born-Infeld black hole metric and reduces to the Reissner-Nordstrom metric in the limit $\beta \rightarrow \infty$ \cite{kim2021}. For $q=0$, it reduces to the metric on the equator of a magnetar \cite{kim2022}. Note that the non-diagonal element ${\tilde g}_{03}$ appears when both the electric charge and the magnetic dipole moment are nonzero. As mentioned before this term appears because the Poynting vector is nonzero. This metric form is quite similar to the metric of the Kerr-Newman black hole. 

Because the observed universe is charge neutral and no charged black hole has ever been observed, we cannot compare the relative strength of the charge term with the mass term. However we can compare the relative strength of the dipole term with the mass term from the observed data on magnetars. Let us estimate the order-of-magnitude contribution to the metric function by the mass term and the magnetic dipole term. For simplicity we consider ${\tilde g}_{11}$. Taking $q = 0$ and restoring the units, we have
\be
- {\tilde g}_{11} = 1 + \frac{ 2GM }{ c^2 r} - \frac{G}{c^4} \frac{\mu_0} {4 \pi} \frac{ \mu^2 }{ r^4} 
+ \frac{1}{\beta^2} \left (  \frac{\mu_0} {4 \pi}  \frac{ \mu }{r^3} \right )^2 .
\ee
One can replace the magnetic dipole moment with the magnetic induction on the surface ($r = r_s$) of the magnetar using 
$ B_s = \frac{\mu_0} {4 \pi}  \frac{ \mu }{r_s^3}$
\be
- {\tilde g}_{11} = 1 + \frac{ 2GM }{ c^2 r_s}  \frac{r_s}{r}  
- \frac{G}{c^4} \left ( \frac{\mu_0} {4 \pi} \right)^{-1} B_s^2 r_s^2 \left ( \frac{r_s}{r} \right )^4 
+ \frac{B_s^2}{\beta^2} \left( \frac{r_s}{r} \right)^6 .
\ee

Because the estimated maximum magnetic field of magnetar is of the order $10^{11} {\rm T}$, we take $\beta \sim 10^{12} {\rm T} $ and estimate the orders of magnitude. 
Taking the mass and radius as those of a typical neutron star, $ M = 1.4 M_{\rm sun} = 2.8 \times 10^{30 } {\rm kg}$ and $r_s = 10 {\rm km}$,
we have
\be
\frac{ 2GM }{ c^2 r_s}   \sim 0.42, 
\ee
\be
 \frac{G}{c^4} \left ( \frac{\mu_0} {4 \pi} \right)^{-1} B_s^2 r_s^2 \sim 8.2 \times 10^{-8} ,
\ee
\be
 \frac{B_s^2}{\beta^2} \sim 10^{-2}.
\ee
The metric contribution by the magnetic dipole originated from the energy momentum tensor is negligible compared with the mass term. 
However, the metric contribution originated from the nonlinear electromagnetic effect is not negligible when a light ray is passing close to the magnetar. 

\section{Geodesic equation and weak bending angle}

In this section we compute the bending angle of light by an astrophysical compact object with electric charge and magnetic dipole. We consider the trajectory of a light ray passing on the equatorial plane described by the following form of metric \cite{Weinberg,hsiao,ali},
\be 
ds^2 = A(r) dt^2 - B(r) dr^2 - C(r)  d \phi^2  + 2 D(r) dt d \phi ,  \label{metriconequator}
\ee
where
\bea
A(r) &=& 1- \frac{ 2GM} { r}  + \frac{ Gq^2 } {r^2} - \frac{1} { \beta^2 } \frac{ q^2 }{r^4} ,  \label{forma}   \\
B(r) &=&  1 + \frac{ 2GM }{ r} - \frac{G \mu^2 }{2 r^4} - \frac{ 1} { \beta^2} \frac{ q^2 }{ r^4 }
+ \frac{ 1 }{\beta^2} \frac{  \mu^2 }{r^6} ,  \label{formb}  \\
C(r) &=&  r^2 \left ( 1 +\frac{2GM}{ r} - \frac {G q^2} {r^4 } + \frac{ 1} { \beta^2} \frac{ \mu^2} {r^6 } \right ),  \label{formc} \\
D(r) &=&  - \frac{Gq \mu}{r^2} + \frac{ 1 }{ \beta^2}\frac{q \mu }{r^4}.   \label{formd}
\eea

The geodesic equation on the equitorial plane of the magnetic dipole can be described by the effective Lagrangian
\be 
{\cal L}_{\rm eff} = \frac{1}{2} {\tilde g}_{\mu \nu} {\dot x}^\mu  {\dot x}^\nu ,
\ee
where the over dot means the derivative with respect to the affine parameter $p$ which describes the trajectory and $ {\tilde g}_{\mu \nu}$, given by Eqs. (\ref{metriconequator})-(\ref{formd}), is the metric restricted to the equatorial plane. 
Because the metric given by Eq. (\ref{metriconequator}) does not depend on $t$ and $\phi$, there are two constants of motion from the associated Killing vectors.
The conserved quantities corresponding to the energy and the angular momentum are
\be
 \frac {\partial {\cal L}_{\rm eff} } {\partial {\dot t} } =  A {\dot t}  + D {\dot \phi} = \varepsilon ,                  \label{energy}
\ee
 \be
 \frac {\partial {\cal L}_{\rm eff} } {\partial {\dot \phi} } = D {\dot t}  - C {\dot \phi} = - \ell .                      \label{angmom}
\ee
One can take $\varepsilon = 1$ by absorbing the constant into the redefinition of the affine parameter. Then, from Eqs. (\ref{energy}) and (\ref{angmom}), we have
\bea
{\dot t} &=& \frac{C - D \ell}{AC +D^2}, \label{tdot}    \\
{\dot \phi} &=& \frac{D+ A \ell}{AC +D^2}.  \label{phidot}
\eea
Inserting Eqs. (\ref{tdot}) and (\ref{phidot}) into the null geodesic condition $ds^2 =0$,  we have
\be 
{\dot r} = \sqrt{ \frac{C - D \ell - A\ell^2}{B(AC +D^2 )}  }.  \label{rdot}
\ee

\begin{figure}
\begin{center}
\includegraphics[height=8cm,keepaspectratio]{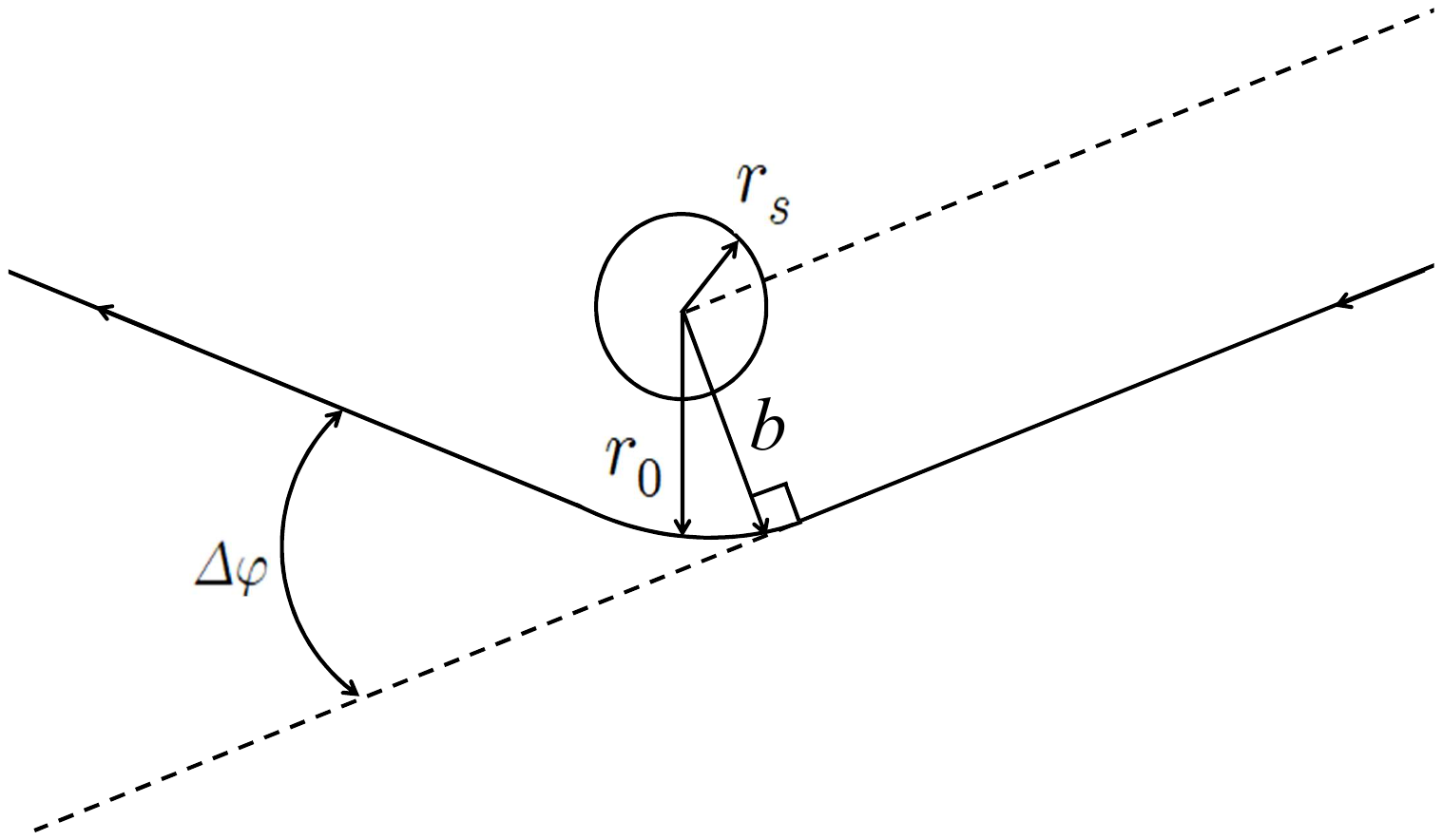}
\caption{Bending angle $\Delta \varphi$ of light ray on the equatorial plane of the magnetic dipole. The distance of closest approach $r_0$ is the minimum of the radial coordinate $r$, $b$ is the impact parameter and $r_s$ is the radius of the compact object with magnetic dipole.} \label{fig1}
\end{center}
\end{figure}

On the equatorial plane, one can find the bending angle (Fig. 1) from \cite{Weinberg}
\be
\Delta \varphi = 2 | \varphi (r_0 ) - \varphi_\infty | - \pi, 
 \label{weinberg855}
\ee
where $r_0$ is the distance of closest approach and 
\be
 \varphi (r_0 ) -  \varphi_\infty = \int_{r_0 }^\infty \frac {d \phi} {d r} dr 
= \int_{r_0 }^\infty 
\frac{\sqrt{B} (D+ A \ell )} {\sqrt{ (AC + D^2) ( C - D \ell - A \ell^2 ) } } dr .
 \label{integral}
 \ee             
Because ${\dot r} = 0$ at the distance of closest approach, the angular momentum can be represented in terms of $r_0$ as
\be
\ell = \frac{ -D_0 + \sqrt{ D_0^2 + 4 A_0 C_0 } }{ 2 A_0} .   \label{ellexact}
\ee

We will compute the integral perturbatively up the first order in $G$ and $1/\beta^2$ in the weak field region. 
Because $D_0^2$ in Eq. (\ref{ellexact}) is second order, we can write $\ell$ as
\be
\ell = \sqrt{ \frac{C_0}{A_0}}  \left( 1 - \frac{D_0} {2 \sqrt{A_0 C_0} } \right ) .   \label{ellfirst}
\ee
Neglecting $D^2$ term in the square root of Eq. (\ref{integral}), we can write
\be
 \frac {d \phi} {d r}
= \sqrt{ \frac{B} {C}} \frac {1 +  \frac{D}{\ell} }  {\sqrt{  \frac{1 }{\ell^2} \frac{C}{A} -1 - \frac{1 }{\ell} \frac{D}{A} } } .
 \label{dphidr1}
 \ee   
One can compute the integral by substituting Eq. (\ref{ellfirst}) into Eq. (\ref{dphidr1}). The details of the integration is given in Appendix B. 
Inserting the result of the integration given by Eq. (\ref{integfinal}) into Eq. (\ref{weinberg855}), we obtain the bending angle up to the first order as
\be
\Delta \varphi =  \frac{4GM}{r_0} - \frac{3 \pi}{4} \frac{ G q^2 }{ r_0^2 } 
    + \frac{\pi}{4}  \frac{Gq \mu}{r_0^3}- \frac{3 \pi}{32} \frac{G \mu^2}{ r_0^4}   
 + \frac{9 \pi}{16}   \frac{q^2}{\beta^2 r_0^4} -  \frac{9 \pi}{16}  \frac{q \mu}{ \beta^2 r_0^5} 
+ \frac{15 \pi}{16}  \frac{\mu^2}{ \beta^2 r_0^6} .    \label{bendang} 
\ee

The bending angle in Eq. (\ref{bendang}) is expressed in terms of the distance of closest approach $r_0$ which is a coordinate dependent measure, while the impact parameter is an invariant. In general relativity coordinates are completely arbitrary, it is important to relate the bending angle to physically measurable quantities \cite{bodennerwill}. One can express the bending angle in terms of the impact parameter $b$ which is coordinate  independent. Because the impact parameter is defined in terms of two conservative quantities of energy and angular momentum, $b = \ell / \varepsilon$, one can relate $b$ to $r_0$ up to the first order, from  Eq. (\ref{ellfirst}),
\be
b =  r_0 \left ( 1 + \frac{2GM}{r_0} - \frac{ G q^2 }{ r_0^2 }  +  \frac{Gq \mu}{2 r_0^3}
+  \frac{q^2}{2 \beta^2 r_0^4} -  \frac{q \mu}{ \beta^2 r_0^5} + \frac{\mu^2}{2 \beta^2 r_0^6} \right ) .    \label{bandr0} 
\ee
Because $b= r_0$ to the leading order, we can replace $r_0$ in Eq. (\ref{bendang}) with $b$ 
\be
\Delta \varphi =  \frac{4GM}{b} - \frac{3 \pi}{4} \frac{ G q^2 }{ b^2 } 
    + \frac{\pi}{4}  \frac{Gq \mu}{b^3}- \frac{3 \pi}{32} \frac{G \mu^2}{b^4}   
 + \frac{9 \pi}{16}   \frac{q^2}{\beta^2 b^4} -  \frac{9 \pi}{16}  \frac{q \mu}{ \beta^2 b^5} 
+ \frac{15 \pi}{16}  \frac{\mu^2}{ \beta^2 b^6} .    \label{bendangitob} 
\ee

The first four terms having the gravitational constant $G$ come from the general relativistic effects while the last three terms having the Born-Infeld parameter $1/\beta^2$ reflect the nonlinear electrodynamic effects. Note that the signs of $Gq^2$ and $G \mu^2$ terms are negative while the signs of $q^2 /\beta^2$ and 
$ \mu^2 /\beta^2$ terms are positive. 
This means that the charge and the magnetic dipole terms originated from $T_{\mu \nu}$ give repulsive effects while those originated  from nonlinear electrodynamic effects give attractive effects. 
The result includes particular cases of weak bending angle in the literature. 
For $q=\mu = 0$, it corresponds to the bending angle by a Schwarzschild black hole. For $\mu = 0$, it corresponds to the bending angle by an Einstein-Born-Infeld black hole \cite{kim2021}. For $\mu = 0$ and $\beta = \infty$, it corresponds to the bending angle by a Reissner-Nordstrom black hole.  
 For $q = 0$, it corresponds to the bending angle of a light ray passing on the equator of a magnetar \cite{kim2022}. 

To compare the results in this nonlinear electrodynamic formalism with those from Einstein-Maxwell formalism, we apply the bending angle to magnetar for the order-of-magnitude estimation. We assume that the magnetar mentioned in Sec. II also have an electric charge $q$. In Reissner-Nordstrom solution, the electric  charge is constrained by the condition \cite{mtw} 
\be
\frac{q^2} {4 \pi \epsilon_0} \le G M^2 .    \label{constraint}
\ee
Parametrizing the fraction of charge $\xi$ as 
\be
q = \sqrt{ {4 \pi \epsilon_0} G M } \xi ,  ~(0 \le \xi \le 1 ) ,  \label{constraint}
\ee
and using the relation between the magnetic dipole moment and the magnetic field on the equator $\mu = (\mu_0 / 4 \pi )^{-1} B_s r_s^3$, we express the bending angle, restoring the units, as 
\bea
\Delta \varphi &=&  \frac{4GM}{c^2 b} - \frac{3 \pi}{4} \frac{ G^2 M^2 \xi^2 }{c^4 b^2 } 
 + \frac{\pi}{4} \frac{{G}^{3 \over 2} M \xi } {c^4 }  \left( \frac{ \mu_0}{4 \pi} \right )^{ - {1 \over 2}} \frac{B_s r_s^3}{b^3}
- \frac{3 \pi}{32} \frac{G}{c^4}  \left( \frac{ \mu_0}{4 \pi} \right )^{ -1} \frac{ B_s^2 r_s^6 }{b^4}   \nonumber   \\  
&+& \frac{9 \pi}{16}  \frac {G }{ 4 \pi \epsilon_0} \frac { M^2 \xi^2} { \beta_E^2} \frac{1}{b^4} 
-  \frac{9 \pi}{16}  \left (  \frac {G}{ 4 \pi \epsilon_0}  \right )^{ 1 \over 2} \frac{M \xi} {\beta_E} \frac{B_s}{\beta_B } \frac{r_s^3 }{b^5}
+ \frac{15 \pi}{16}  \frac{B_s^2}{ \beta_B^2} \frac{r_s^6}{ b^6} ,      \label{bendingest} 
\eea
where $\beta_{E(B)}$ is the electric (magnetic) Born-Infeld parameter ($\beta_E = c \beta_B$). 

In Reissner-Nordstrom solution, the parameter $\xi$ can be any value between zero and one ($0\le \xi \le 1$). However in Born-Infeld gravity, it cannot be arbitrary because the possible electric field should be smaller than the electric Born-Infeld parameter $\beta_E$. If this condition is not satisfied, the effective metric obtained from Eq. (\ref{gtilde}) is not valid. This gives stricter constraint for the parameter $\xi$. Assuning a uniform distribution of charge, we can roughly estimate the upper bound of $\xi$ from the condition that the electric field on the equator of the magnetar should be below $\beta_E$
\be 
\frac{1}{4 \pi \epsilon_0} \frac{q}{r_s^2} =\sqrt{  \frac{G}{4 \pi \epsilon_0} }  \frac{M \xi }{r_s^2} < \beta_E = c \beta_B . \label{xiconstraint}
\ee
For $M= 1.4 M_{\rm sun}$, $r_s = 10 {\rm km}$ and $\beta_B = 10^{12} {\rm T}$, this gives $ \xi < 1.38 \times 10^{-2}$. Thus we estimate the possible maximum bending angle for $\xi=10^{-3}$ and $B_s = 10^{11} {\rm T}$. The maximum bending angle can be obtaind at
$b = r_s = 10 {\rm km}$ corresponding to a light ray glancing the equator of the magnetar. 
Note that the Schwarzschild radius for $M= 1.4 M_{\rm sun}$ is $ r_{\rm Sch} = 4.15 {\rm km }$ and accordingly the photon sphere is 
$r_{\rm ps} = 6.22 {\rm km} $. We define the relative deviation  from the Schwarzschild bending angle $\Delta \varphi_{\rm Sch}  = 4 GM /c^2 b (= 8.30 \times 10^{-1}  {\rm rad}$) as
\be
\delta \equiv \frac{  \Delta \varphi - \Delta \varphi_{\rm Sch} }{  \Delta \varphi_{\rm Sch} }.   \label {relativedeviation}
\ee
Taking the limit $ \beta \rightarrow \infty$, we obtain the relative bending angles in Einstein-Maxwell gravity

i-a) RN black hole ($B_s = 0$):  $\delta = - 1.22 \times 10^{-7} $,

i-b) magnetar ($\xi =0$):   $\delta = -2.92 \times 10^{-8} $,

i-c) charged magnetar: $\delta = -1.31 \times 10^{-7} $,

\noindent and taking $\beta_B = 10^{12} {\rm T}$, we obtain the relative bending angles in Einstein-Born-Infeld gravity 

ii-a) BI black hole ($B_s = 0$): $\delta = 1.11 \times 10^{-2}$,

ii-b) BI magnetar ($\xi =0$): $\delta = 3.55 \times 10^{-2} $,

ii-c) BI charged magnetar: $\delta = 3.12 \times 10^{-2} $.

\noindent The relative bending angles in Einstein-Maxwell theory are negligibly small compared with those in  Einstein-Born-Infeld gravity.
Because the magnetic field of magnetar is the strongest known in the universe, observations of the lensed image properties produced by magnetar can provide a means of testing magnetic dipole solutions in alternative gravity theories if the allowed precision is sufficient enough. 

\section{Conclusion}

We consider the geometrical aspects of light around a compact astrophysical object with both the electric field and the magnetic field in Einstein-Born-Infeld theory. 
We consider the null geodesic of a light ray when it passes a massive object with electric charge and magnetic dipole. The effective metric can be obtained from the light cone condition reflecting the nonlinear electromagnetic effects. 
We found the asymptotic form of the effective metric up to the first order in gravitational constant $G$ and Born-Infeld parameter $1/\beta^2$ on the equatorial plane. Then we compute the bending angle of light from the geodesic equation. 
We confirm that the result covers particular cases in the literature. 

In this work we compute the effective metric from Eq. (\ref{g_eff}) which does not show birefringence. This is possible only in the classical Born-Infeld theory where the effective action depends only one parameter $\beta$. In general the effective nonlinear action can be polarization-dependent where the parameters of the invariants $S$ and $P$ are different \cite{Kruglov10}. For example, it is well-known that the effective metric of Euler-Heisenberg Lagrangian is polarization-dependent \cite{HeisenbergEuler}. One can find the polarization-dependent effective metric following the formalism by Novello et al \cite{Novello}. 

In this study we compute the bending angle for a special case where a light ray is passing on the equatorial plane of the magnetic dipole. It will be interesting to find the bending angle for an arbitrary orientation of the magnetic dipole. For magnetars considered in this paper, the Schwarzschild radius and the photon sphere are smaller than the radius of magnetar. However,  for charged black holes with magnetic dipole, the effective metric used in this paper might be useful to study the strong lensing.

\section*{Acknowledgements}
This work was supported by Basic Science Research Program through the National Research Foundation of Korea (NRF) funded by the Ministry of Education, Science and Technology (NRF-2019R1F1A1060409).

\appendix

\section{~}

 In this appendix we compute the effective metric from Eq. (\ref{gtilde}). 
The contravariant tensor of the metric given by Eq. (\ref{metricftn}) is 
\be
g^{\mu \nu} = \frac{1}{s} \pmatrix {r^2 e^{-\nu}  & 0  & 0  & \omega \cr
0 & -s e^{-\lambda}  &  0 & 0 \cr  
0&  0& -s e^{-\lambda} &0  \cr
\omega&  0& 0  & -  e^{\rho}  } ,  \label{metriccontra}
\ee
where $ s = r^2 e^{\rho - \nu} + \omega^2$. 
We will compute the effective metric up to the first order in $G$ and $1/\beta^2$.
Because the nonlinear electrodynamic correction in Eq. (\ref{gtilde}) is of the order $1/ \beta^2$, we can use the electromagnetic tensor in flat space given by Eq. (\ref{Fco}). The contravariant form of the field tensor is 
\be
F^{\mu \nu} 
 = \pmatrix {0  & \Phi_r & \Phi_z  & 0 \cr 
-\Phi_r & 0  &  0 & -\frac{\psi_r}{r^2} \cr  
-\Phi_z&  0& 0 &-\frac{\psi_z}{r^2}  \cr
0& \frac{ \psi_r}{r^2}& \frac{\psi_z}{r^2} & 0  } .  \label{Fcontra}
\ee
Using Eqs. (\ref{metriccontra}) and  (\ref{Fcontra}), the nonzero components of $\delta g^{\mu\nu}$are
\bea
\delta g^{00}&=&  \frac{1}{ \beta^2} ( \Phi_r^2 + \Phi_z^2 ) e^\lambda ,   \\
\delta g^{03} &=& \delta g^{30}  = \frac{1}{\beta^2} \left ( \frac{\Phi_r \psi_r } { r^2 } + \frac{\Phi_z \psi_z}{ r^2} \right ) e^\lambda , \\
\delta g^{11} &=&  -\frac{ 1}{ \beta^2} \left ( \Phi_r^2 e^\rho + 2 \frac{\Phi_r \psi_r } { r^2} \omega - \frac{\psi_r^2} {r^2 } e^{- \nu} \right ) , \\
\delta  g ^{12} &=& \delta  g ^{21} =- \frac{1}  {\beta^{2}} \left ( \Phi _{r} \Phi  _{z} e ^{\rho } + \frac {\Phi_{r} \psi_{z} 
+ \psi_{r}\ \Phi_{z}} {r ^{2}} \omega  - \frac{\psi_{r} \psi_{z}} {r^{2}} e ^{- \nu } \right ) , \\
\delta g^{22} &=& - \frac{1} { \beta^2 } \left ( \Phi_z^2 e^\rho + \frac {2 \Phi_z \psi_z  } { r^2 }\omega 
- \frac{ \psi_z^2 }{ r^2} e^{-\nu}  \right ) , \\
\delta  g ^{33} &=& \frac {1} {\beta^{2}} \left ( \frac{\Phi_{r}^{2}} {r^{2}} + \frac{\psi_{z}^{2}} {r ^{2}} \right) e^{\lambda }.
\eea

The elplicit form of the effective metric can be obtained upon substituting Eqs. (\ref{erho})-(\ref{psi}) to $\delta g^{\mu \nu}$. 
Because we are interested in computing the bending angle on the equatorial plane ($z=0$), the nonzero comopnent of the effective metric up to 
the first order in $G$ and $1/\beta^2$ are
\bea
{\tilde g}^{00} &=& 1+ \frac{ 2GM} { r}  - \frac{ Gq^2 } {r^2}  + \frac{ 1 } { \beta^2 } \frac{ q^2 }{r^4} ,    \\
{\tilde g}^{03} &=& {\tilde g}^{30} = - \frac{Gq \mu}{r^4} + \frac{ 1 }{ \beta^2}\frac{q \mu }{r^6 },   \\
{\tilde g}^{11} &=& - \left ( 1 - \frac{ 2GM }{ r} + \frac{G \mu^2 }{2 r^4} + \frac{ 1} { \beta^2} \frac{ q^2 }{ r^4 }
- \frac{ 1 }{\beta^2} \frac{  \mu^2 }{r^6} \right ) ,   \\
{\tilde g}^{22} &=& - \left ( 1 - \frac{ 2GM }{ r }+ \frac{G \mu^2}{2 r^4 } \right ) , \\
{\tilde g}^{33} &=& - \frac{ 1} { r^2} \left ( 1 -\frac{  2GM}{ r} + \frac {G q^2} {r^4 } - \frac{ 1} { \beta^2} \frac{ \mu^2} {r^6 } \right ). 
\eea

\section{~}
In this appendix we compute the integral in Eq. (\ref{integral}) up to the first order in $G$ and $1/\beta^2$. 
Because $D_0$ is first order in  Eq. (\ref{ellfirst}), substituting $D_0 / \sqrt{A_0 C_0} = D_0 /r_0$  to the leading order, we have
\be
\ell = \sqrt{ \frac{C_0}{A_0}}  \left( 1 - \frac{D_0} {2 r_0} \right ) .  \label{ellfirstorder}
\ee
Inserting this, Eq. (\ref{dphidr1}) can be written as 
\be
 \frac {d \phi} {d r}
= \sqrt{ \frac{B} {C}} \frac {1 +  \frac{D}{r_0}  }  
{  \sqrt{ \frac{C}{A} \frac{A_0}{C_0}  + \frac{r^2}{r_0^2} \frac{D_0} {r_0}  -1  -  \frac{D}{r_0 }   } }.
 \ee   
Using the explicit forms of metric functions in Eqs. (\ref{formb}) and (\ref{formc}), we have
\be 
\sqrt{ \frac{B}{C} } = \frac{x}{r_0} \left( 1 + \frac{Gq^2}{2 r_0^2} x^2 -  \frac{G \mu^2}{4 r_0^4} x^4  
             -  \frac{q^2}{2 \beta^2 r_0^4} x^4 \right ) ,                \label{sqrtboverc1}
\ee
where $x$ is defined as $x \equiv r_0 / r$. 
Similarly, the numerator and the denominator can be written as
\be
1 + \frac{D}{r_0} = 1 - \frac{Gq \mu}{r_0^3} x^2 +  \frac{q \mu} {\beta^2 r_0^5} x^4 ,   \label{numerator1}
\ee
\bea
\frac{C}{A} \frac{A_0}{C_0} + \frac{r^2}{r_0^2} \frac{D_0} {r_0} - 1 -  \frac{D}{r_0 } 
&=& \frac{1- x^2}{x^2} \bigg [ 1 - \frac{4GM}{r_0} \frac{1} {1 + x} + \frac{2 G q^2 }{r_0^2 } - \frac{q^2}{\beta^2 r_0^4} (1 + x^2 )     \nonumber  \\
&-& \frac{\mu^2}{\beta^2 r_0^6} (1 + x^2 + x^4 )                     
-  \frac{Gq \mu}{ r_0^3} (1 + x^2 ) +  \frac{q \mu}{\beta^2 r_0^5} (1 + x^2 + x^4)  \bigg ] . \label{sqrt1}
\eea
Substituting Eqs. (\ref{sqrtboverc1}) - (\ref{sqrt1}) and $dr = - (r_0 / x^2 ) d x$, the integral can be written as
\bea
\int_{r_0 }^\infty \frac {d \phi} {d r} dr  
&=& \int_{0 }^1 \frac{ d x} { \sqrt{ 1 - x^2}} \bigg [  1 + \frac{2GM}{r_0} \frac{1} {1 + x} + \frac{ G q^2 }{2 r_0^2 } (x^2 -2) 
    + \frac{Gq \mu}{2r_0^3}(1 - x^2 ) - \frac{G \mu^2}{4 r_0^4} x^4         \nonumber    \\
&+&   \frac{q^2}{2\beta^2 r_0^4} (1 + x^2 - x^4 )   -  \frac{q \mu}{2 \beta^2 r_0^5} (1 + x^2 - x^4) 
+  \frac{\mu^2}{2 \beta^2 r_0^6} (1 + x^2 + x^4 )     \bigg ] .
\eea
Upon integration we finally have 
\be
\int_{r_0 }^\infty \frac {d \phi} {d r} dr   = \frac{\pi}{2} + \frac{2GM}{r_0} - \frac{3 \pi}{8} \frac{ G q^2 }{ r_0^2 } 
    + \frac{\pi}{8}  \frac{Gq \mu}{r_0^3}- \frac{3 \pi}{64} \frac{G \mu^2}{ r_0^4}       \nonumber    \\
 + \frac{9 \pi}{32}   \frac{q^2}{\beta^2 r_0^4} -  \frac{9 \pi}{32}  \frac{q \mu}{ \beta^2 r_0^5} 
+ \frac{15 \pi}{32}  \frac{\mu^2}{ \beta^2 r_0^6} .    \label{integfinal} 
\ee

\end{document}